\documentclass[onecolumn,showpacs,preprintnumbers,amsmath,amssymb]{revtex4}

\usepackage{graphicx}
\usepackage{dcolumn}
\usepackage{bm}

\begin{document}


\title{Disk illumination by black hole superradiance of electromagnetic perturbations}

\author{Taichi Kobayashi}
 \email{kobayashi@gravity.phys.nagoya-u.ac.jp}
\author{Kohei Onda}
 \email{onda@gravity.phys.nagoya-u.ac.jp}
\author{Akira Tomimatsu}%
 \email{atomi@gravity.phys.nagoya-u.ac.jp}
\affiliation{
Department of Physics, Nagoya University, Chikusa-Ku, Nagoya, 464-8602, Japan
}%


\date{\today}

\begin{abstract}
Using the Kerr-Schild formalism to solve the Einstein-Maxwell equations,
 we study energy transport due to time-dependent electromagnetic
 perturbations 
 around a Kerr black hole,
 which may work as a mechanism to illuminate a disk located on the
 equatorial plane.
 For such a disk-hole system it is found that the energy extraction 
 from the hole can occur under the well-known superradiance condition 
 for wave frequency, even though the energy absorption into the hole
 should be rather dominant near the polar region of the horizon.
 We estimate the efficiency of the superradiant amplification of the
 disk illumination.
 Further we calculate the time-averaged energy density distribution to
 show explicitly the existence of a negative energy region near the
 horizon
 and to discuss the possible generation of a hot spot on the disk.
\end{abstract}

\maketitle

\section{Introduction}

It is widely believed that there exists a rotating black hole surrounded
by a disk in the central region of highly energetic astrophysical
objects,
such as active galactic nuclei (AGNs), X-ray binary systems, 
and gamma-ray bursts(GRBs).
In particular, 
magnetic fields in the disk-black hole system are expected to play an
important role in the processes of disk radiation, jet production, 
and so on.
As was emphasized in \cite{BH-D1,BH-D2},
if a black hole is magnetically connected with a disk, 
the energy and angular momentum fluxes can be transported between them
along the magnetic field lines through a mechanism analogous to the
Blandford-Znajek effect \cite{BZ}.
The energy supply to an accretion disk due to spin-down of a rapidly
rotating black hole will enhance the disk radiation \cite{MC1,MC2},
which may be relevant to the observation of extremely broad and
red-shifted
Fe K$\alpha$ line emission from a nearby Seyfert 1 galaxy \cite{MCObs}.

Though it is possible to construct stationary magnetospheric models
representing the magnetic connection 
(e.g., see \cite{SBH-DV,SBH-DF,KBH-DF} 
for vacuum and force-free models with disk currents),
the stability of such a configuration is not confirmed.
In fact, recent numerical simulations of general relativistic
magnetohydrodynamics (GRMHD) rather claim that the magnetic connection
should be disrupted to produce open field lines threading the event
horizon and extending to infinity \cite{McKinney1,McKinney2,Koide2}.
Such a change of configuration of magnetic field lines can develop
turbulent disturbances in the inner magnetospheric region, 
and the Poynting flux of strongly disturbed electromagnetic fields may
propagate toward the equatorial plane to illuminate the disk surface.
The subsequent dissipation of the supplied electromagnetic energy inside
the disk should contribute to a disk heating.
Hence, in this paper, as a possible mechanism of energy transport to the
disk we would like to pay our attention to the process of the disk
illumination caused by persistent excitation of electromagnetic
disturbances in the disk-black hole system.

In general, the injected electromagnetic disturbances should be
scattered away to infinity, or absorbed by a black hole.
Here we are not interested in the scattered outgoing part.
We rather consider absorption by a thin disk, which corresponds to the
boundary condition that the energy flux is transported to the equatorial
plane both from the upper side and from the lower one.
This boundary condition of the thin disk means the existence of
discontinuities of components of electromagnetic fields (namely, the
existence of surface currents) 
on the equatorial plane.
Of course, the exact analysis based on GRMHD is required to study the
time evolution of electromagnetic fields in black hole magnetospheres. 
In this paper, nevertheless, we focus on the analysis of the vacuum
Maxwell equations under the disk boundary condition as a first step to
approach the problem of disk illumination.
This is because our main purpose is to reveal the superradiance effect
in Kerr background geometry which transports the energy from a central
black hole to ``a surrounding disk''.
If there exists no disk, the efficiency of superradiance of vacuum
electromagnetic waves to amplify the energy radiated away to infinity is
well-known \cite{SR}.
We expect that such an amplification can also occur in the process of
disk illumination, by which a hot spot may appear on the disk surface
near the inner edge.

To treat vacuum electromagnetic fields in the disk-black hole system, it
is convenient to use the Kerr-Schild formalism \cite{DKS} for solving
the Einstein-Maxwell equation.
If this formalism is applied to obtain electromagnetic perturbations on
Kerr background, it is known that all the field components are simply
derived by two arbitrary complex functions $\psi$ and $\phi$ \cite{B2},
which can be appropriately chosen according to the disk boundary
condition.
In Sec.~\ref{sec:KS}, we briefly review the derivation of
electromagnetic fields around a Kerr black hole in the framework of the
Kerr-Schild formalism. 
Unfortunately, 
the time-dependent electromagnetic solutions obtained here 
can describe only the part of incident waves 
(namely, any waves scattered toward infinity are absent),
if the regularity on the event horizon is required.
This will correspond to the case that the disk illumination and the
black hole absorption become dominant for the injection of
electromagnetic disturbance. In Sec.~\ref{sec:model}, we assume a
relation between the two arbitrary complex functions $\psi$ and $\phi$
to determine the electromagnetic disturbances, which is useful to see
the superradiance effect.
Then the field components are expressed by using the Boyer-Lindquist
coordinate system, from which in Sec.~\ref{sec:GP},
the spatial distribution of the time-averaged Poynting flux is given.
Further, it is shown how the total energy fluxes illuminating the disk
surface and extracted from the black hole are dependent on the wave
frequency of electromagnetic disturbance. 
We find that near the polar
region of the event horizon the electromagnetic energy is always
supplied to the black hole irrespective of the wave frequency, in
contrast to the usual superradiant scattering. It is confirmed, however,
that the energy transport from the black hole to the disk can occur, if
the wave frequency satisfies the usual superradiance condition.
The maximal superradiant amplification for the disk illumination is
roughly estimated.
In Sec.~\ref{sec:SE}, the complex functions $\psi$ and $\phi$ are
specified to evaluate numerically the Poynting flux and the energy
density of electromagnetic perturbations. 
The existence of a hot spot is claimed in terms of the distributions of
the energy density deposited in the disk.
The extraction of the angular momentum from the black hole is also
discussed in Appendix \ref{sec:app}.
Hereafter
we use units such that $c=G=1$.

\section{Electromagnetic fields in the Kerr-Schild formalism\label{sec:KS}}

Let us consider electromagnetic fields
in the framework of the Kerr-Schild formalism (see details in \cite{DKS}).
Though 
this formalism is introduced to solve the full Einstein-Maxwell equations,
it may be applied to obtain electromagnetic perturbations on
Kerr background.
The metrical ansatz is 
\begin{equation}
 g^{\mu\nu}
  =
  \eta^{\mu\nu}-2He^{3\mu}e^{3\nu},
  \label{eq:KSform}
\end{equation}
where $\eta^{\mu\nu}$ is the metric of an auxiliary Minkowski spacetime, 
$H$ is scalar function,
and
$e^{3\mu}$ is a null vector field, which is tangent to a geodesic
and shear-free principal null congruence.
It is convenient to calculate the Einstein-Maxwell equations 
using tetrad components.
All other null tetrad
vectors are defined by the condition
\begin{equation}
 g_{ab}=e_{a}^{\ \mu}e_{b\mu}=
  \left(
   \begin{array}{cccc}
    0 & 1 & 0 & 0 \\
    1 & 0 & 0 & 0 \\
    0 & 0 & 0 & 1 \\
    0 & 0 & 1 & 0
   \end{array}
  \right)
  =g^{ab},
\end{equation}
where latin and greek suffixes mean tetrad  and tensor suffixes, respectively. 
A tensor $T_{\mu\ ...}^{\ \nu ...}$
is related to its tetrad components $T_{a\ ...}^{\ b ...}$
by either of the two equivalent relations
\begin{equation}
 T_{a\ ...}^{\ b...}=e_{a}^{\mu}e^{b}_{\nu}\ ...\ T_{\mu\ ...}^{\ \nu...},
  \
  T_{\mu\ ...}^{\ \nu ...}=e^{a}_{\mu}e_{b}^{\nu}\ ...\ T_{a\ ...}^{\ b...}.
\end{equation}

The essential point of the Kerr-Schild formalism is 
to use the complex form of electromagnetic field tensors
given by
\begin{equation}
 {\mathcal F}_{\mu\nu}
  \equiv F_{\mu\nu}+\frac{1}{2}i\epsilon_{\mu\nu\rho\sigma}F^{\rho\sigma},
\end{equation}
where $\epsilon_{\mu\nu\rho\sigma}$ is completely skew-symmetric and
equal to $(-g)^{\frac{1}{2}}$ when $\mu\nu\rho\sigma=1234$.
The corresponding null tetrad components are
\begin{equation}
 {\mathcal F}_{ab}=F_{ab}+\frac{1}{2}i\epsilon_{abcd}F^{cd},
  \label{eq:emtet}
\end{equation}
where $\epsilon_{abcd}$ is completely skew-symmetric and
$\epsilon_{1234}=i$.
By virtue of the difinition (\ref{eq:emtet}) and the Einstein equations,
the tetrad component ${\cal F}_{32}$, ${\cal F}_{41}$, and 
${\cal F}_{42}$ are found to be zero.
The electromagnetic fields are completely determined by only two complex
components ${\cal F}_{12}$, and ${\cal F}_{31}$. 
It is interesting to note that the Kerr-Schild form remain valid, even
if a back reaction on the gravitational field by the electromagnetic
field is considered.

A part of the Maxwell equations 
allows to write the tetrad components as
\begin{eqnarray}
 {\mathcal F}_{12}={\mathcal F}_{34}&=&AZ^{2},
  \label{eq:F1tet}\\
 {\mathcal F}_{31}&=&\gamma Z -(AZ)_{,1},
  \label{eq:F2tet}
\end{eqnarray}
where $Z$ is the complex expansion of 
the null vector $e^3$, and 
commas denote the directional derivatives along chosen
null tetrad vectors.
The functions $A$ and $\gamma$ should be
determined by solving the other Maxwell equations.

For $\gamma=0$,
the stationary solutions including the Kerr-Newman solution are
obtained \cite{DKS}.
However,
for $\gamma\neq 0$, it is difficult to obtain the exact solutions of
the Einstein-Maxwell equations.
Therefore, we restrict our consideration to electromagnetic
perturbations on Kerr background.
In the Kerr-Schild coordinates, the Kerr metric is given by
\begin{eqnarray}
 {\mathrm d}s^2
  &=&
  -{\mathrm d}\tilde{t}^2
  +{\mathrm d}r^2
  +\Sigma {\mathrm d}\theta^2 
  +(r^2+a^2)\sin^2\theta {\mathrm d}\tilde{\varphi}^2
  \nonumber\\ & & 
  -2a\sin^2\theta {\mathrm d}r {\mathrm d}\tilde{\varphi}
  +\frac{2Mr}{\Sigma}({\mathrm d}\tilde{t}+{\mathrm d}r-a
  \sin^2\theta{\mathrm d}\tilde{\varphi})^2,
\end{eqnarray}
where $\Sigma\equiv r^2+a^2\cos^2\theta$, and $M$ and $a$ denote the
mass and the angular momentum per unit mass of the black hole, respectively.
The function $H$ in Eq.~(\ref{eq:KSform}) is given by $2Mr/P^2\Sigma$,
where $P=1/\sqrt{2}\cos^2(\theta/2)$,
and the null tetrad vectors are given by
\begin{subequations}
\label{allequations}
\begin{eqnarray}
 e^{1}
  &=&
  2^{-\frac{1}{2}}e^{i\tilde{\varphi}}
  \bigglb[
   \tan(\theta/2){\mathrm d}\tilde{t}
   +\tan(\theta/2){\mathrm d}r
   +(r+ia\cos\theta){\mathrm d}\theta
   -(a-ir)\sin\theta {\mathrm d}\tilde{\varphi}
  \bigglb],\\
 e^{2}
  &=&
  2^{-\frac{1}{2}}e^{-i\tilde{\varphi}}
  \bigglb[
   \tan(\theta/2){\mathrm d}\tilde{t}
   +\tan(\theta/2){\mathrm d}r
   +(r-ia\cos\theta){\mathrm d}\theta
   -(a+ir)\sin\theta {\mathrm d}\tilde{\varphi}
  \bigglb],\\
 e^{3}
  &=&
  P
  (
   {\mathrm d}\tilde{t}+{\mathrm d}r-a\sin^2\theta {\mathrm d}\tilde{\varphi}
  ),\\
 e^{4}
  &=&
  2^{-\frac{1}{2}}
  (
  -{\mathrm d}\tilde{t}+\cos\theta {\mathrm d}r -r\sin\theta {\mathrm d}\theta
  )
  +
  He^{3},
\end{eqnarray}
\end{subequations}
where $e^3$ is defined as an ingoing null geodesic.
Then, for electromagnetic perturbations
on Kerr background.
We obtain $A$ and $\gamma$ described as \cite{B2}
\begin{eqnarray}
 A&=&\frac{\psi(Y,\tau)}{P^2}, \\
 \gamma&=&\frac{2^{1/2}\psi_{,\tau}}{P^2Y}+\frac{\phi(Y,\tau)}{P},
\end{eqnarray}
where 
$\tau=\tilde{t}+r+ia\cos\theta$ 
because $e^{3}$ is chosen as an ingoing vector field, and 
$Y=e^{i\tilde{\varphi}}\tan(\theta/2)$.
Now the expansion $Z$ in Eqs.~(\ref{eq:F1tet}) and (\ref{eq:F2tet}) is
given by $Z/P=1/(r+ia\cos\theta)$. 
It should be noted that tetrad components include the two arbitrary
complex functions $\psi(Y,\tau)$ and $\phi(Y.\tau)$, and are written by
\begin{widetext}
 \begin{eqnarray}
  {\mathcal F}_{12}
   &=&
   \frac{\psi}{(r+ia\cos\theta)^2},
   \label{eq:f12}\\
  {\mathcal F}_{31}
   &=&
   \frac{1}{r+ia\cos\theta}
   \bigglb\{
    e^{-i\tilde{\varphi}}
    \left[
     \frac{2\cos^2(\theta/2)}{\tan(\theta/2)}
     \frac{r+ia}{r+ia\cos\theta}\psi_{,\tau}
     +\sin\theta\frac{r-ia}{(r+ia\cos\theta)^2}\psi
    \right]
    \nonumber \\ &&
    +\phi
    -\frac{\psi_{,Y}}{r+ia\cos\theta}
   \bigglb\}.\label{eq:f31}
 \end{eqnarray}
\end{widetext}

It is well-known that any complex function which is not a constant
cannot be regular everywhere on the complex plane.
In the following section we will assume the existence of a branch cut in
$\phi$ and $\psi$ placed on the complex $Y$-plane, which corresponds to
the existence of a disk current on the equatorial plane $\theta=\pi/2$.

\section{Superradiant disturbances with disk currents\label{sec:model}}

In this section,
we give some constraints to $\psi$ and $\phi$, which will be useful to
see clearly the superradiant energy transport in the disk-black hole system.
Further, following the usual analysis of energy extraction from a black
hole,
we introduce the Boyer-Lindquist coordinates, which lead to the metric
\begin{eqnarray}
 {\mathrm d}s^2
  &=&-\left(1-\frac{2Mr}{\Sigma}\right){\mathrm d}t^2
  -\frac{4aMr}{\Sigma}
  \sin^2\theta {\mathrm d}t {\mathrm d}\varphi
  +\frac{\Sigma}{\Delta}{\mathrm d}r^2
  +\Sigma {\mathrm d}\theta^2
  +\frac{{\mathcal A}}{\Sigma}\sin^2\theta {\mathrm d}\varphi^2,
\end{eqnarray}
where
$\Delta=r^2+a^2-2Mr$, and 
${\mathcal A}=(r^2+a^2)^2-a^2\Delta\sin^2\theta$.
The Boyer-Lindquist coordinates $t$ and $\varphi$ 
are related to Kerr-Schild coordinates $\tilde{t}$ and $\tilde{\varphi}$
as follows,
\begin{eqnarray}
 {\mathrm d}t&=& 
  {\mathrm d}\tilde{t}-\frac{2Mr}{\Delta}{\mathrm d}r,\\
 {\mathrm d}\varphi&=&
  {\mathrm d}\tilde{\varphi}-\frac{a}{\Delta}{\mathrm d}r.
  \label{eq:psico}
\end{eqnarray}

Recall that superradiant modes with the frequency $\omega$ have the form 
$f(r,\theta)e^{i(m\varphi-\omega t)}$,
if no disk boundary exists.
This motivates us to assume that the perturbations given by
Eqs.~(\ref{eq:f12}) and (\ref{eq:f31}) correspond to the $m=1$ modes,
and $\psi$ is written by the form
\begin{equation}
 \psi(Y,\tau)
 \equiv \psi(X),\ 
 X\equiv e^{-i\omega\tau}Y=e^{-i\omega\tau+i\tilde{\varphi}}\tan(\theta/2),
\end{equation}
where 
$\tilde{t}$ in $\tau$ and $\tilde{\varphi}$ are the Kerr-Schild
coordinates which remains finite even on the horizon.

To impose the similar constraint on $\phi$, let us consider an energy
flux vector ${\cal E}^{\mu}$ defined in the Boyer-Lindquist frame as 
\begin{equation}
 {\mathcal E}^{\mu}\equiv -T^{\mu}_{ \ t},
  \label{eq:energyf}
\end{equation}
where $T_{\mu\nu}$ is the stress-energy tensor of electromagnetic field
written by
\begin{equation}
 T_{\mu\nu}=\frac{1}{4\pi}
  \left(
   F_{\mu\gamma}F_{\nu}^{\ \gamma}
   -\frac{1}{4}g_{\mu\nu}F_{\alpha\beta}F^{\alpha\beta}
   \right).
\end{equation}
Using Eqs.~(\ref{eq:f12}) and (\ref{eq:f31}),
we can obtain the radial component ${\mathcal E}^r$ as 
\begin{eqnarray}
 {\mathcal E}^{r}
  &=&
  -\frac{1}{16\pi\Sigma}
  \left|
  \frac{2}{\sin\theta}
  \bigglb\{
  Y\phi
   -\frac{X\psi_{,X}}{r+ia\cos\theta}
  \right.
  \left[1+2i\omega(r+ia)\cos^2(\theta/2)\right]
  \bigglb\}
  \left.
  +\frac{ia\sin\theta}{(r+ia\cos\theta)^2}\psi
  \right|^2
  \nonumber \\ &&
  +\frac{a^2\sin^2\theta}{16\pi\Sigma^3}|\psi|^2
  .\label{eq:erg}
\end{eqnarray}
It is clear that the energy extraction from the hole 
(namely, ${\cal E}^r>0$ on the horizon $r=r_{\textrm H}$)
becomes possible at the region where the second term of the right-hand
side of Eq.~(\ref{eq:erg}) is dominant.
Hence, it is difficult to extract the energy near the polar region where
$\sin\theta$ is very small. 
Here we consider the case that ${\cal E}^r$ on the horizon becomes zero
at the equator $\theta=\pi/2$, which leads to the requirement
\begin{equation}
 \phi(Y,\tau)=e^{-i\omega\tau}
  \left[
   (r_{\textrm H}+ia)i\omega +1
  \right]
  \frac{\psi_{,X}}{r_{\textrm H}},
  \label{eq:phicond}
\end{equation}
This means that the energy extraction may be more efficient at an
intermediate region between the pole and the equator, producing a hot
spot on the disk surface slightly apart from the horizon through the
propagation of the Poynting flux.

Under the condition (\ref{eq:phicond}) we obtain the electromagnetic
component written in Boyer-Lindquist coordinate system as follows
\begin{widetext}
\begin{subequations}
\label{allequations}
\begin{eqnarray}
 F_{tr}&=&{\rm Re}
 \left[
 \frac{\psi}{(r+ia\cos\theta)^2}
 +\frac{a}{\Delta}\frac{iX\psi_{,X}}{2r_{\textrm H}(r+ia\cos\theta)}K(r,\theta)
 \right],\\
 F_{t\theta}&=&{\rm Re}
 \left[
 -\frac{ia\sin\theta\psi}{(r+ia\cos\theta)^2}
 +\frac{X\psi_{,X}}{r_{\textrm H}(r+ia\cos\theta)\sin\theta}
 K(r,\theta)
 \right],\label{eq:Etheta}\\
 F_{t\varphi}&=&{\rm Re}
 \left[
 \frac{iX\psi_{,X}}{2r_{\textrm H}(r+ia\cos\theta)}
 K(r,\theta)
 \right],
 \\
 F_{\theta\varphi}&=&
 {\rm Re}
 \left[
 -\frac{i(r^2+a^2)\sin\theta \psi}{(r+ia\cos\theta)^2}
 +\frac{a\sin\theta X\psi_{,X}}{r_{\textrm H}(r+ia\cos\theta)}
 K(r,\theta)
 \right],
 \label{eq:Br}
 \\
 F_{r\varphi}&=&{\rm Re}
 \left[
 \frac{a\sin^2\theta \psi}{(r+ia\cos\theta)^2}
 +\frac{r^2+a^2}{\Delta}
 \frac{iX\psi_{,X}}{2r_{\textrm H}(r+ia\cos\theta)}
 K(r,\theta)
 \right],\\
 F_{r\theta}
  &=&
  \frac{\Sigma}{\Delta\sin\theta}{\rm Re}
 \left[\frac{X\psi_{,X}}{r_{\textrm H}(r+ia\cos\theta)} 
 K(r,\theta)
 \right],\label{eq:Bphi}
\end{eqnarray}
\label{eq:eminB}
\end{subequations}
\end{widetext}
where $K(r,\theta)\equiv  (r-r_{\textrm
H})(1-a\omega)+ia\cos\theta\left(1-\omega/\Omega(r)\right)$, 
 $\Omega(r)\equiv a/(r r_{\textrm H}+a^2)$,
and then 
$\Omega_{\textrm H}\equiv \Omega(r_{\textrm H})$ is 
the angular velocity of the black hole.
It is easy to check that 
the electromagnetic invariant
$F_{\mu\nu}F^{\mu\nu}$ is finite on the horizon.
As was previously mentioned, there should be a singularity in $\psi(X)$
on the complex X-plane, which in this paper is assumed to be due to the
existence of disk currents on the equator.
This means that the components $F_{t\theta}$, $F_{\theta\varphi}$ and 
$F_{r\theta}$ (namely, the imaginary part of $\psi$ and the real part of
$X\psi_{,X}$) become discontinuous at $\theta=\pi/2$. Such a
discontinuity will be generated if a branch point in $\psi$ exists at
$X=e^{i\beta}$ where $\beta$ is a real constant (see the example given
in Sec.~\ref{sec:SE}).
Note that the absolute value $|X|=e^{a\omega\cos\theta}\tan(\theta/2)$
becomes equal to unity at $\theta=\pi/2$. 
The branch point $X=e^{i\beta}$ may appear on some conical plane
$\theta=\theta_{0}$($\neq \pi/2$) giving $|X|=1$ if $a\omega>1$.
Hence, in the following, the allowed range of the frequency $\omega$ is
limited to $0<\omega<1/a$, for which we obtain $|X|<1$ in the upper
region $0<\theta\leq\pi/2$ and $|X|>1$ in the lower region
$\pi/2<\theta\leq \pi$.

We must also consider the regularity condition for $F_{\mu\nu}$ at the
polar axis (i.e., at $\theta=0$, $\pi$).
Noting that $|X|\simeq \sin(\theta/2)$ in the limit $\theta\to 0$
and $|X|\simeq 1/\cos(\theta/2)$ in the limit $\theta \to \pi$,
we find the boundary condition for $\psi$ to be
\begin{equation}
 \psi(X)\to X^2,\label{eq:psirest1}
\end{equation}
for $\theta \to 0$, and
\begin{equation}
 \psi(X)\to \frac{1}{X^2},\label{eq:psirest2}
\end{equation}
for $\theta \to \pi$.

\section{Distribution of electromagnetic energy flux\label{sec:GP}}

Now let us discuss the energy flux distribution given by the
electromagnetic fields with the component (\ref{eq:eminB}).
Our special attention will be paid to the amplitudes at infinity, on the
horizon and on the disk surface, and the amplification of the disk
illumination due to the superradiance effect will be clarified.
For this purpose we calculate the energy flux vector defined by 
Eq.~(\ref{eq:energyf})
and obtain
\begin{subequations}
\label{allequations}
\begin{eqnarray}
 {\mathcal E}^t
  &=&
  -\frac{r^2+a^2}{\Delta}{\mathcal E}^r
  \nonumber\\ & &
  +\frac{a}{4\pi r_{\textrm H}\Sigma^3}
  {\rm Re}\left[
     iX\psi_{,X}\bar{\psi}
      K(r,\theta)
      (r+ia\cos\theta)
    \right]
  +\frac{2a^2\sin^2\theta+\Sigma}{8\pi\Sigma^3}|\psi|^2,
  \\
 {\mathcal E}^r
  &=&
  -\frac{1}{16\pi\Sigma^2}
   \left|
    \frac{
    K(r,\theta)
    }
    {r_{\textrm H}\sin\theta}
    2X\psi_{,X}
    -i\frac{a\sin\theta}{r+ia\cos\theta}\psi
   \right|^2
   +\frac{a^2\sin^2\theta}{16\pi\Sigma^3}
   \left|
    \psi
   \right|^2
   ,
  \\
 {\mathcal E}^{\theta}
  &=&
  \frac{1}{4\pi\Sigma^3r_{\textrm H}\sin\theta}
  {\rm Re}
  \left[
   X\psi_{,X}\bar{\psi}
   K(r,\theta)
   (r+ia\cos\theta)
  \right],\\
 {\mathcal E}^{\varphi}
  &=&
  -\frac{a}{\Delta}{\mathcal E}^{r}
  +\frac{1}{4\pi r_{\textrm H}\Sigma^3\sin^2\theta}
  {\rm Re}\left[
     iX\psi_{,X}\bar{\psi}
     K(r,\theta)
     (r+ia\cos\theta)
    \right]
  +\frac{a}{4\pi\Sigma^3}|\psi|^2,
\end{eqnarray}
\end{subequations}
where bar denotes complex conjugate.
It is easy to check that the conservation law 
${\cal E}^{\mu}_{\ ;\mu}=0$ is satisfied.

Note that the complex variable $X$ in $\psi$ is oscillatory with respect
to the Kerr-Schild time $\tilde{t}$ (as well as the azimuthal angle
$\tilde{\varphi}$).
Then, the energy flux vector ${\cal E}^{\mu}$ contains oscillatory
terms.
To estimate a real efficiency of the energy transport, we must consider
the time-averaged quantities such that
\begin{equation}
 \langle A\rangle \equiv\frac{\omega}{2\pi}\int_{0}^{2\pi/\omega} A d\tilde{t}.
\end{equation}
Furthermore, we can obtain useful results for the time-averaged
quantities,
by expanding $\psi$ as follows
\begin{equation}
 \psi=\sum_{n}a_{n}X^{n},
\end{equation}
where $n$ runs from $1$ to $\infty$ for $0<|X|<1$ (corresponding to the
upper region $0<\theta<\pi/2$),
while it runs from $-1$ to $-\infty$ for $1<|X|<\infty$ (corresponding
to the lower region $\pi/2<\theta<\pi$).
Such an expansion is possible, because $\psi(X)$ is assumed to be
regular except at branch points on the equatorial plane $|X|=1$.
By using the expansion form, for example, we obtain
\begin{eqnarray}
 X\psi_{,X}\bar{\psi}
  =
  \sum_{n}\sum_{m}
  n a_{n}\bar{a}_{m}X^{n}\bar{X}^{m},
\end{eqnarray}
for which the time average leads to
\begin{eqnarray}
 \langle X\psi_{,X}\bar{\psi}\rangle
  =\sum_{n}
  n
  |a_{n}|^2|X|^{2n}.\label{eq:taxpsix}
  \label{eq:expxpsix}
\end{eqnarray}
Because $\langle X\psi_{,X}\bar{\psi}\rangle$ turns out to be real as
well as $\langle |\psi|^2\rangle$ and $\langle|X\psi_{,X}|^2\rangle$,
the time-averaged quantities of ${\cal E}^{\mu}$ are given by
\begin{widetext}
\begin{subequations}
 \label{allequations}
\begin{eqnarray}
 \langle {\mathcal E}^{t} \rangle
  &=&
  -\frac{r^2+a^2}{\Delta}\langle {\mathcal E}^{r} \rangle
  -\frac{a^2\cos\theta}{4\pi \Sigma^3 r_{\textrm H}}
  \langle X\psi_{,X}\bar{\psi}\rangle
  \left[
  (r-r_{\textrm H})(1-a\omega)+r(1-\omega/\Omega(r))
  \right]
  \nonumber \\ &&
  +\frac{2a^2\sin^2\theta+\Sigma}{8\pi\Sigma^3}
  \langle |\psi|^2 \rangle,\label{eq:et}
  \\
 \langle {\mathcal E}^{r} \rangle
  &=&
  -\frac{1}{4\pi \Sigma^2 r_{\textrm H}^{\ 2}\sin^2\theta}
  \langle |X\psi_{,X}|^2\rangle
  \left[
   (r-r_{\textrm H})^2(1-a\omega)^2+a^2\cos^2\theta(1-\omega/\Omega(r))^2
  \right]
  \nonumber \\
 &&
  +\frac{a^2\cos\theta}{4\pi\Sigma^3r_{\textrm H}}
  \langle X\psi_{,X}\bar{\psi}\rangle
  \left[
   (r-r_{\textrm H})(1-a\omega)+r(1-\omega/\Omega(r))
  \right],
  \label{eq:er}
  \\
 \langle {\mathcal E}^{\theta} \rangle 
  &=&
  \frac{1}{4\pi\Sigma^3 r_{\textrm H}\sin\theta}
  \langle X\psi_{,X}\bar{\psi}\rangle
  \left[
   r(r-r_{\textrm H})(1-a\omega)-a^2\cos^2\theta (1-\omega/\Omega(r))
  \right],
  \label{eq:etheta}\\
 \langle {\mathcal E}^{\varphi} \rangle
  &=&
  -\frac{a}{\Delta}\langle {\mathcal E}^{r} \rangle
  -\frac{a\cos\theta}{4\pi r_{\textrm H}\Sigma^3\sin^2\theta}
  \langle X\psi_{,X}\bar{\psi}\rangle
  \left[
   (r-r_{\textrm H})(1-a\omega)+r(1-\omega/\Omega(r))
  \right]
  \nonumber \\ &&
  +\frac{a}{4\pi\Sigma^3}\langle |\psi|^2\rangle.
\end{eqnarray}
\label{eq:energyf}
\end{subequations}
\end{widetext}

In Kerr-Schild formalism considered here, the perturbations at infinity
are purely incident waves without any scattered outgoing waves.
From Eq.~(\ref{eq:er}) it is easily seen that the amplitude 
$-\langle {\cal E}^{r}\rangle_{\infty}$ of the incident energy flux
per unit area at infinity is given by
\begin{equation}
 -\langle {\mathcal E}^{r} \rangle_{\infty}
  =
  \frac{\langle |X\psi_{,X}|^2\rangle}{4\pi r_{\textrm H}^{\ 2}\sin^2\theta}
  \frac{(1-a\omega)^2+\omega^2 r_{\textrm H}^{\ 2}\cos^2\theta}{r^2}.
\end{equation}
Of course, the total incident flux $E_{\infty}$ remains finite, which
is written by
\begin{equation}
 E_{\infty}= \frac{(1-a\omega)^2}{2r_{\rm H}^{\ 2}}
  \int_{0}^{\pi}
  \frac{\langle |X\psi_{,X}|^2\rangle}{\sin\theta}
  [(1-a\omega)^2+\omega^2r_{\textrm H}^{\ 2}\cos^2\theta]
  {\mathrm d}\theta.
\end{equation}

On the other hand the outgoing energy flux 
$\langle {\cal E}^r\rangle_{\textrm H}$ per unit area on the horizon may
be positive by virtue of the superradiance effect.
In fact, from Eq.~(\ref{eq:er}) we have
\begin{eqnarray}
 \langle{\mathcal E}^{r}\rangle_{\textrm H}
 & = &
 -\frac{a^2\cos\theta (1-\omega/\Omega_{\textrm H})}{4\pi
 \Sigma_{\textrm H}^{\ 2}r_{\textrm H}^{\ 2}\sin^{2}\theta}
 \bigglb[
  \langle |X\psi_{,X}|^2\rangle
  \cos\theta (1-\omega/\Omega_{\textrm H})
  -\frac{r_{\textrm H}^{\ 2}\sin^2\theta}{\Sigma_{\textrm H}}
  \langle X\psi_{,X}\bar{\psi}\rangle
 \bigglb]\label{eq:erh}.
\end{eqnarray}
Note that from Eq.~(\ref{eq:expxpsix}) the time-averaged quantity 
$\langle X\psi_{,X}\bar{\psi}\rangle$ becomes positive in the region
where $\cos\theta>0$, while it becomes negative in the region where 
$\cos\theta<0$.
Then, it is assured that $\langle {\cal E}^r\rangle_{\textrm H}$ 
remains negative
for $\omega>\Omega_{\textrm H}$, namely, the electromagnetic energy is
supplied to the black hole.
If we consider the low frequency range $0<\omega<\Omega_{\textrm H}$,
it is easy to find that $\langle {\cal E}^r\rangle_{\textrm H}$ becomes
positive near the equator where $|\cos\theta|\ll 1$, even though it is
still negative near the polar region where $\sin\theta\ll 1$.
We can evaluate the net flux $E_{\textrm H}$ integrated over the whole
region of the horizon, by using the expansion form (\ref{eq:expxpsix}), 
and the result is
\begin{eqnarray}
 E_{\textrm H}
  &=&
  \iint\langle{\mathcal E}^{r}\rangle_{\textrm H}\Sigma_{\textrm H}\sin\theta
  {\mathrm d}\theta {\mathrm d}\varphi 
  \nonumber \\
 &=&
  \frac{a\omega(1-\omega/\Omega_{\textrm H})}{2r_{\textrm H}^{\ 2}}
   \int_{0}^{\infty}
  \frac{{\mathrm d}|X|}{|X|}\langle|X\psi_{,X}|^2\rangle
  \frac{\cos^2\theta}{1-a\omega\sin^2\theta}.
  \label{eq:teh}
\end{eqnarray}
We can verify that the energy extraction from the black hole occurs for
incident wave with the frequency $\omega$ in the range
$0<\omega<\Omega_{\textrm H}$ in accordance with the result of the usual
superradiant scattering \cite{SR}.
It should be emphasized that the condition for $E_{\textrm H}>0$ does
not depend on the details of $\psi$.
Of course, some suitable choice of $\psi$ becomes necessary to maximize
the value of $E_{\textrm H}$.

Finally we turn our attention to the energy flux vector ${\cal E}^\mu$
on the disk surface to confirm that the extracted energy is transported
for the disk illumination.
Noting that $\sqrt{\Sigma}=r$ at $\theta=\pi/2$,
the energy flux ${\cal E}_{\textrm D}$ per unit area injected to the
disk can be evaluated as
\begin{equation}
 {\cal E}_{\textrm D}(r)=r\langle{\cal E}^{\theta}\rangle_{+}
  +r\langle{\cal E}^{\theta}\rangle_{-},
  \label{eq:efd}
\end{equation}
where from Eq.~(\ref{eq:etheta}) we obtain
\begin{equation}
 r\langle {\cal E}^{\theta}\rangle_{\pm}
  =\pm \frac{1}{4\pi r_{\textrm H}}
  \frac{(r-r_{\textrm H})(1-a\omega)}{r^4}
  \langle X\psi_{,X}\bar{\psi}\rangle_{\pm},
  \label{eq:retheta}
\end{equation}
corresponding to the injection to the upper and lower disk
surfaces, respectively.
The factor $\langle X\psi_{,X}\bar{\psi}\rangle_{\pm}$
in Eq.~(\ref{eq:retheta}) may be calculated from Eq.~(\ref{eq:expxpsix})
in the limit $|X|\to 1$ with positive and negative integers $n$,
respectively.
The energy flux ${\cal E}_{\textrm D}$ to illuminate the disk is found
to become maximum at the position $r=4r_{\rm H}/3$.
Further, it is easy to obtain the total energy flux $E_{\textrm D}$ as follows
\begin{eqnarray}
 E_{\textrm D}&=&
  2\pi \int_{r_{\textrm H}}^{\infty}{\cal E}_{\textrm D}r{\mathrm d}r
  \nonumber\\
  &=&\frac{1-a\omega}{4r_{\textrm H}^2}
  \left[
   \langle X\psi_{,X}\bar{\psi}\rangle_{+}
   -\langle X\psi_{,X}\bar{\psi}\rangle_{-}
  \right].
\end{eqnarray}
Because we can show the equality
\begin{equation}
 \int_{0}^{\infty}\frac{{\mathrm d}|X|}{|X|}\langle
  |X\psi_{,X}|^2\rangle
  =\frac{1}{2}
  \left[
   \langle X\psi_{,X}\bar{\psi}\rangle_{+}
   -\langle X\psi_{,X}\bar{\psi}\rangle_{-}
  \right],
  \label{eq:relation}
\end{equation}
by the help of the expansion form of $\psi$, we have the relation
\begin{equation}
 E_{\textrm D}=E_{\infty}+E_{\textrm H},
  \label{eq:econs}
\end{equation}
which is a result of the conservation law ${\cal E}^{\mu}_{\ ;\mu}=0$.
The ratio $E_{\textrm H}/E_{\infty}$($=E_{\textrm D}/E_{\infty}-1$)
can represent an efficiency of the energy extraction from the hole
(namely, the amplification of the disk illumination)
due to the superradiance.
Note that the factor $\cos^2\theta/(1-a\omega\sin^2\theta)$
in Eq.~(\ref{eq:teh}) is smaller than unity.
Then, by virtue of Eq.~(\ref{eq:relation}) we have the upper bound such
that 
\begin{equation}
 E_{\textrm H}/E_{\textrm D}
  <\frac{a\omega(1-\omega/\Omega_{\textrm H})}{1-a\omega}.
  \label{eq:upb}
\end{equation}
If we consider the external black hole with $a=M$, the right-hand side
of Eq.~(\ref{eq:upb}) becomes maximum at 
$\omega/\Omega_{\textrm H}=2-\sqrt{2}$, and we arrive at the result
\begin{equation}
 E_{\textrm H}/E_{\infty}<(\sqrt{2}-1)/2.
  \label{eq:ub}
\end{equation}
In the next section we will calculate the ratio 
$E_{\textrm H}/E_{\infty}$ by specifying the complex function $\psi$.
It is also interesting to note that the disk illumination is quite
suppressed
if no superradiant contribution exists
(namely, for the case $\omega>\Omega_{\textrm H}$).
In fact, in the limit $\omega\to 1/a$ the energy flux $E_{\textrm D}$
on the disk surface approaches zero,
and the energy flux $E_{\infty}$ incident from infinity is almost
absorbed by the central black hole.

As an additional comment let us discuss the energy density on the disk
surface,
which is derived by the use of Eq.~(\ref{eq:et}) in the limit 
$\theta\to \pi/2$.
Because 
the time-averaged value $\langle|X\psi_{,X}|^2\rangle_{+}$ and 
$\langle|\psi|^2\rangle_{+}$ evaluated on the upper disk surface
may be different from $\langle|X\psi_{,X}|^2\rangle_{-}$ and 
$\langle|\psi|^2\rangle_{-}$, respectively, evaluated on the lower one,  
we consider the mean energy density defined by
\begin{equation}
  \langle {\cal E}^{t}\rangle_{\textrm D}
 =\frac{1}{4\pi}
  \bigglb[
   \frac{(r^2+a^2)(r-r_{\textrm H})^2(1-a\omega)^2}{\Delta r^4
   r_{\textrm H}^{\ 2}}\langle|X\psi_{,X}|^2\rangle_{\textrm D}
   +\frac{2a^2+r^2}{2r^6}\langle|\psi|^2\rangle_{\textrm D}
  \bigglb],
  \label{eq:edens}
\end{equation}
where 
$\langle |X\psi_{,X}|^2\rangle_{\textrm D}
=(\langle|X\psi_{,X}|^2\rangle_{+}+\langle|X\psi_{,X}|\rangle_{-})/2$
and $\langle|\psi|^2\rangle_{\textrm D}
=(\langle |\psi|^2\rangle_{+}+\langle |\psi|^2\rangle_{-})/2$.
This describes the energy density distribution in the disk as a function
of $r$, which may have a maximum at the position slightly apart from the
horizon.
The energy density deposited in the disk may dissipate, for example, via
a Joule heating.
Then, a hot spot will appear on the disk surface near the maximum point
of the energy density $\langle{\cal E}^{t}\rangle_{\textrm D}$.
It is known that a region which negative energy density can exist near
the horizon if superradiant scattering occurs.
The distribution of the electromagnetic energy density in the whole
region around the disk-black hole system will be illustrated in the next
section, aiming to show the location of such a hot spot region and to
verify the existence of the negative energy region.

\section{A specific example\label{sec:SE}}

In the previous section we obtained the time-averaged energy flux vector
$\langle {\cal E}^{\mu}\rangle$ written by an arbitrary complex function 
$\psi(X)$, and we pointed out some essential features of the
superradiant energy transport in the disk-black hole system.
To see more clearly such an energy transport process, in this section,
we calculate numerically $\langle {\cal E}^{\mu}\rangle$ under a
specific choice of $\psi(X)$.
Considering the regularity conditions (\ref{eq:psirest1}) 
and (\ref{eq:psirest2}) given at the polar axis,
the complex function $\psi(X)$ is assumed to be
\begin{equation}
 \psi(X)
  =
  \psi_{0}
  \left[
   (X^{-2}+X^2)^{3/2}-X^{-3}-X^{3}
  \right]^2,
  \label{eq:SExamp}
\end{equation}
where $\psi_{0}$ is a real constant.
For this choice of $\psi$,
there exist four branch points at $X^2=\pm i$, and the derivative
$\psi_{,X}$ can remain finite there.
By virtue of the existence of the branch points,
the imaginary part of $\psi$ and the real part of $X\psi_{,X}$ become
discontinuous at $\theta=\pi/2$.
This leads to 
$\langle X\psi_{,X}\bar{\psi}\rangle_{+}
=-\langle X\psi_{,X}\bar{\psi}\rangle_{-}\simeq 76.629\psi_{0}^{\ 2}$
in Eq.~(\ref{eq:retheta}) giving the energy flux illuminating the disk surface.
Because this form of $\psi$ is invariant under the transformation
$X\to 1/X$,
the time-averaged values $\langle|\psi|^2\rangle$
and $\langle|X\psi_{,X}|^2\rangle$ can remain continuous even at 
$\theta=\pi/2$, and we obtain 
$\langle|\psi|^2\rangle_{\textrm D}\simeq 18.862\psi_{0}^{\ 2}$
and $\langle|X\psi_{,X}|^2\rangle_{\textrm D}\simeq 359.60\psi_{0}^{\ 2}$
in Eq.~(\ref{eq:edens}) giving the energy density deposited in the disk.

\begin{figure}
\includegraphics[width=245pt]{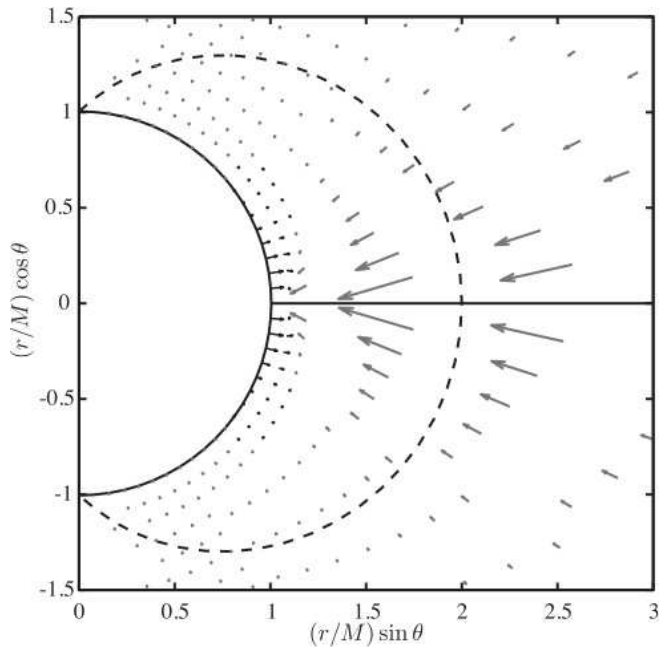}
\caption{\label{fig:energy}
 Time-averaged energy transport in the disk-black hole system.
 The electromagnetic disturbance is given by Eq.~(\ref{eq:SExamp})
 with the wave frequency $\omega\simeq 0.76238\Omega_{\rm H}$,
 and the spin parameter a is chosen as $a=0.99999M$.
 The arrows show the poloidal energy flux with the components 
 $\langle{\cal E}^r\rangle$ and 
 $\sqrt{\Delta}\langle{\cal E}^\theta\rangle$, which are normalized by
 $\psi_{0}^{\ 2}/M^4$.
 The dark gray and light gray arrows correspond to the fluxes with 
 $\langle {\cal E}^r\rangle>0$
 and $\langle{\cal E}^r\rangle<0$, respectively.
 The horizon and the equatorial disk are shown as solid lines, and the
 boundary of the ergosphere is shown as a dashed line.
}
\end{figure}
As the first application of $\psi$ given by Eq.~(\ref{eq:SExamp}) the
time-averaged poloidal energy flow is shown in Fig.~\ref{fig:energy}.
The radial and zenithal components measured in an orthonormal basis may
be given by $\sqrt{\Sigma/\Delta}\langle{\cal E}^r\rangle$ and
$\sqrt{\Sigma}\langle{\cal E}^\theta\rangle$, respectively.
In Fig.~\ref{fig:energy}, however, the poloidal components expressed by
the arrows correspond to $\langle{\cal E}^r\rangle$ and
$\sqrt{\Delta}\langle{\cal E}^\theta\rangle$.
This allows us to see the finite energy flux 
$\langle {\cal E}^r\rangle_{\textrm H}$ on the horizon.
We note that see the finite energy flux
$\langle{\cal E}^r\rangle_{\textrm H}\geq 0$ on the horizon is limited
to the range $-\theta_{\textrm m}\leq \theta-\pi/2\leq \theta_{\textrm m}$ where 
$\theta_{\textrm m}\simeq0.69743$.
It is easy in Fig.~\ref{fig:energy} to see that the energy flux from the
horizon in the range $-\theta_{\textrm m}< \theta-\pi/2<\theta_{\textrm m}$ is transported to
the disk surface. 

\begin{figure}
\includegraphics[width=245pt]{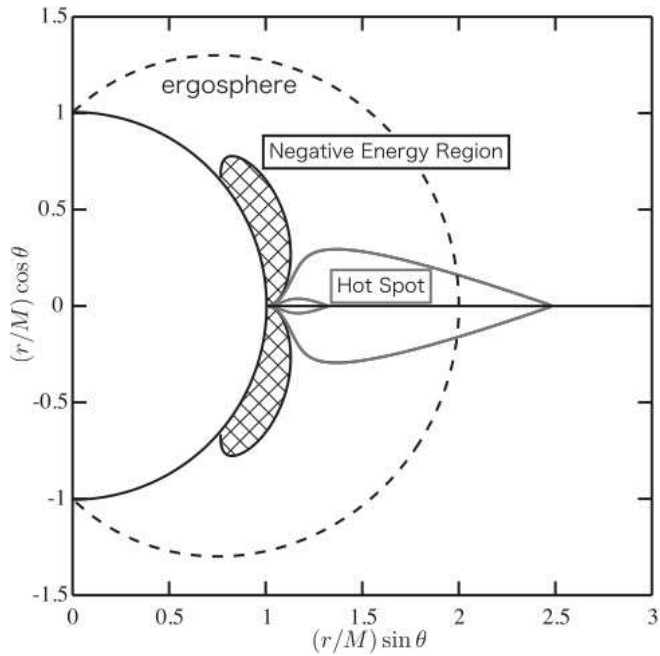}
\caption{\label{fig:density}
Contour of time-averaged energy density for the electromagnetic
 disturbance given by Eq.~(\ref{eq:SExamp}) on the
 $(r/M)\sin\theta$-$(r/M)\cos\theta$ plane.
 The wave frequency and the spin parameter are assumed to be
 $\omega\simeq 0.76238\Omega_{\rm H}$ and $a=0.99999M$
 The negative energy region is shown as the shaded region.
 The position of the maximum energy density $\langle{\cal E}^{t}\rangle_{\textrm m}$ is 
 at $r\simeq 1.0535M$.
 The surface giving 
 $\langle{\cal E}^{t}\rangle=\langle{\cal E}^{t}\rangle_{\textrm m}/2$ 
 (which may be explained as the
 boundary of the hot spot) is shown by the short gray line.
 To see the rapid decrease of $\langle{\cal E}^{t}\rangle$, the surface giving
 $\langle{\cal E}^{t}\rangle=\langle{\cal E}^{t}\rangle_{\textrm m}/10$ 
 is also shown by the long gray line.
 The horizon and the equatorial disk are shown as solid line, 
 and the boundary of the ergosphere is shown as a dashed line.
}
\end{figure}
Next, let us consider the spatial distribution of the time-averaged
energy density $\langle{\cal E}^t\rangle$ derived by the specific choice
of $\psi(X)$.
As will be seen Figs.~\ref{fig:energy} and \ref{fig:density}, the region
with $\langle{\cal E}^{t}\rangle<0$ is found to appear inside the region
with $\langle{\cal E}^r\rangle>0$.
It is also interesting to note that the energy density 
$\langle{\cal E}^{t}\rangle$ becomes maximum on the disk at the radius
slightly apart from the horizon (i.e., at $r\simeq 1.0488r_{\rm H}$).
The region with a high energy density such that 
$\langle{\cal E}^{t}\rangle\geq\langle{\cal E}^t\rangle_{\textrm m}/2$ (where 
$\langle{\cal E}^{t}\rangle_{\textrm m}$ is the maximum density) is shown in
Fig.~\ref{fig:density}.
If the deposited electromagnetic energy density dissipates to heat up
the disk, a hot spot as expected to appear on the inner part of the
disk including such a high energy density.
As was mentioned in the previous section the energy flux
(\ref{eq:efd})
per unit area to illuminate the disk surface becomes maximum at the
radius $r=4r_{\textrm H}/3$. 
The energy supply is also efficient in the inner part of disk near the
horizon, and may balance the enrgy dissipation to keep the hot spot formation.

\begin{figure}
\includegraphics[width=245pt]{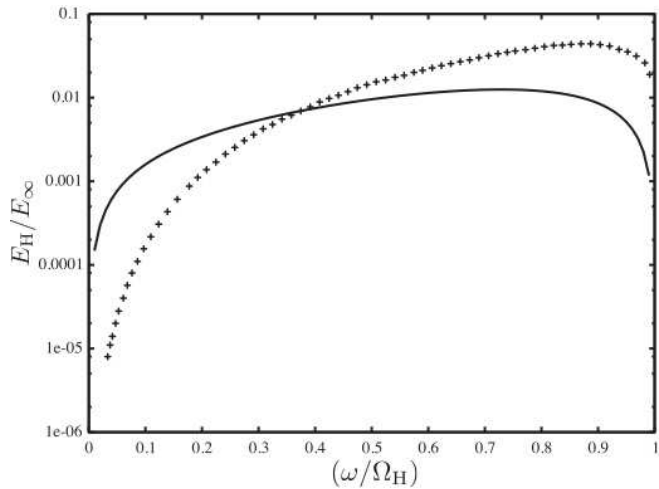}
\caption{\label{fig:hikaku}
 The efficiency $E_{\textrm H}/E_{\infty}$ of energy extraction
 in the disk-black hole system as a function of the wave frequency
 $\omega$
 in the range $0<\omega<\Omega_{\textrm H}$. 
 The spin parameter is fixed as $a=0.99999M$, to compare with 
 the most efficient
 superradiant scattering amplification (corresponding to the mode $m=1$
 and $\ell=1$) of vacuum electromagnetic waves estimated in \cite{SR},
 which is plotted by the points.
 The efficiency obtained from Eq.~(\ref{eq:SExamp}) is shown as a solid line.
 }
\end{figure}
\begin{figure}
\includegraphics[width=245pt]{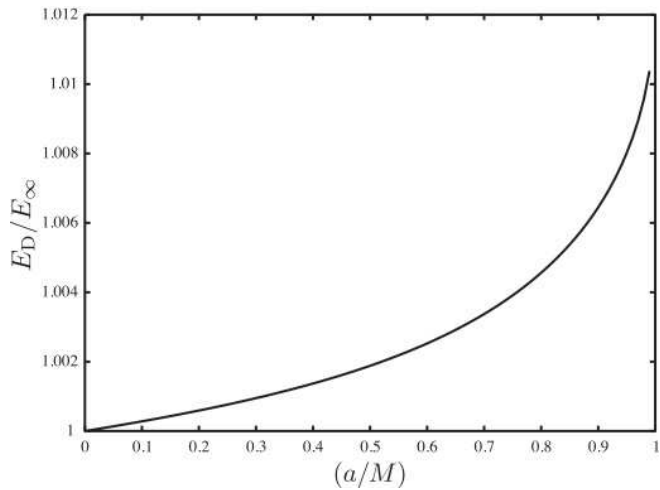}
\caption{\label{fig:hikaku2}
 The efficiency $E_{\textrm D}/E_{\infty}$
 of the disk illumination as a function of the spin parameter $a$.
 The wave frequency for the electromagnetic disturbance given by 
 Eq.~(\ref{eq:SExamp})
 is fixed as $\omega\simeq 0.76238\Omega_{\textrm H}$.}
\end{figure}
Finally let us calculate in more details the efficiency 
$E_{\textrm H}/E_{\infty}$ of the superradiant energy extraction from
the black hole.
In Fig.~\ref{fig:hikaku} the ratio $E_{\textrm H}/E_{\infty}$ is plotted
as a function of the wave frequency $\omega$.
We note that the maximum value of $E_{\textrm H}/E_{\infty}$ is much
smaller than upper bound shown in Eq.~(\ref{eq:ub}). 
It may be possible to choose $\psi(X)$ allowing a more efficent energy
extraction. 
Nevertheless we can claim from Fig.~\ref{fig:hikaku} that the efficiency 
$E_{\textrm H}/E_{\infty}$ obtained have can be larger (or not so
smaller) than the efficiency of superradiant scattering without disk
illumination evaluated in \cite{SR}.
It is sure that even though in the disk-black hole system the energy
flux $\langle{\cal E}^{r}\rangle_{\textrm H}$ per unit area on the
horizon becomes negative near the polar region,
the net flux $E_{\textrm H}$ is not so suppressed.
The energy extraction from the black hole induces the amplification of
the disk illumination according to the conservation law (\ref{eq:econs}).
If the wave frequency is fixed to be $\omega\simeq 0.76238\Omega_{\textrm H}$ 
(corresponding to the maximum efficiency shown in
Fig.~\ref{fig:hikaku},
we can plot the efficiency $E_{\textrm D}/E_{\infty}$ of the disk 
illumination as a function of the spin parameter $a$.
It is clear from Fig.~\ref{fig:hikaku2} that $E_{\textrm D}/E_{\infty}$
increases monotonically as $a$ increases.
If this result does not crucially depend on the specific choice of
$\psi(X)$,
we arrive at the conclusion that electromagnetic disturbance generated
around a
``near-extremal'' black hole can most efficiently amplify the disk illumination.

\appendix
\section{Extraction of angular momentum\label{sec:app}}

Here let us discuss the relation between 
the energy flux and the angular momentum flux on the horizon.
If the rotational energy is extracted from the black hole,
then the angular momentum flux may be also extracted,
which is defined by
\begin{equation}
 {\mathcal L}^{\mu}\equiv T^{\mu}_{\ \varphi}.
\end{equation}
By using the same procedure to obtain 
$\langle{\cal E}^{r}\rangle_{\textrm H}$,
the time-averaged radial angular momentum flux 
is given by
\begin{equation}
 \langle{\mathcal L}^r\rangle_{\textrm H}
  =
  \langle{\mathcal E}^r\rangle_{\textrm H}/\Omega_{\textrm H}
   +\frac{a(1-\omega/\Omega_{\textrm
   H})^2\cos^2\theta}{4\pi r_{\textrm H}\Sigma_{\textrm H}\sin^2\theta}
  \langle|X\psi_{,X}|^2\rangle.
\end{equation}
It is easy to see that 
the angular momentum on the horizon is extracted from the black
hole,
when the energy extraction occurs, i.e., 
$\langle{\mathcal E}^r\rangle_{\textrm H}>0$.
However,
if the net angular momentum flux $L_{\textrm H}$ on the horizon defined as
\begin{equation}
 L_{\textrm H}\equiv
  \iint\langle{\mathcal L}^{r}\rangle_{\textrm H}\Sigma_{\textrm H}\sin\theta
  {\mathrm d}\theta {\mathrm d}\varphi. 
\end{equation}
is evaluated, we can recover the relation
\begin{equation}
 E_{\textrm H}=\omega L_{\textrm H}.
\end{equation}
Then, if the wave frequency $\omega$ is in the range
$0<\omega<\Omega_{\textrm H}$, we obtain $L_{\textrm H}>0$.
The angular momentum can be transported from the black hole to the disk
via electromagnetic disturbances. 

\newpage 
\bibliography{paper}

\end{document}